\documentclass[a4paper,10pt]{article}
\usepackage{amsmath}
\usepackage{amsfonts}
\usepackage{amssymb}
\usepackage{graphicx}
\usepackage{cite}

\begin{document}

\date{}

\title{\bf Two fluid anisotropic dark energy models in a scale invariant theory}

\author{S. K. Tripathy\footnote{Department of Physics,
Indira Gandhi Institute of Technology,
Sarang, Dhenkanal, Odisha-759146, INDIA,
tripathy\_ sunil@rediffmail.com}, B. Mishra\footnote{Department of Mathematics, Birla Institute of Technology and Science-Pilani,
Hyderabad Campus,Hyderabad-500078, INDIA, bivudutta@yahoo.com} and P. K. Sahoo\footnote{Department of Mathematics, Birla Institute of Technology and Science-Pilani,
Hyderabad Campus,Hyderabad-500078, INDIA, sahoomaku@rediffmail.com}}

\maketitle

\begin{abstract}
Some anisotropic Bianchi V dark energy models are investigated in a scale invariant theory of gravity. We consider two non interacting fluids such as dark energy and a bulk viscous fluid. Dark energy pressure is considered to be anisotropic in different spatial directions. A dynamically evolving  pressure anisotropy is obtained from the models.  The models favour phantom behaviour. It is observed that, in presence of dark energy, bulk viscosity has no appreciable affect on the cosmic dynamics.

\end{abstract}


\section{Introduction}

Observations from type Ia Supernova, Baryon Acoustic Oscillations (BAO), X-ray clusters, large scale structure have confirmed the late time cosmic speed up phenomena \cite{Riess98, Allen04, Ein05}. A hypothetical fluid with almost static density and a negative pressure called dark energy is responsible for this phenomena. In theoretical front, dark energy is usually modelled through a cosmological constant. Because of the puzzles like fine tuning and coincidence problem many dynamically evolving dark energy candidates have been proposed. These dark energy candidates dynamically evolve with cosmic time. Alternative geometrical approaches have also been developed to address the cosmic speed up phenomenon \cite{Nojiri03, Harko11, Harko10, Nojiri05, Bamba12}. Even though the dark energy and cosmic acceleration has triggered a good deal of research interest in recent time, dark energy still remains as a mystery. 

Besides the issue of late time cosmic acceleration, the observed anisotropy in temperature power spectrum has triggered much attention in recent times. According to the the standard cosmological model ($\Lambda$CDM model), the universe is mostly flat and spatially isotropic. However, some observable anomaly at large scale \cite{Antoniou12} and observational data from Planck collaboration show that $\Lambda$CDM model does not fit well to the temperature power spectrum at least at low multipoles  \cite{Ade13a}.   In order to address these issues, some anisotropic plane symmetric models have been proposed in recent times \cite{Campa06, Campa09, Gruppo07}. In this context, Bianchi type models (which are the generalization of FRW cosmology) can be suitable to handle the smallness in the angular power spectrum of temperature anisotropy. 

In the context of late time cosmic dynamics dominated by dark energy and the anisotropy in CMB temperature power spectrum, many authors have investigated different models considering anisotropy in dark energy pressure in the framework of Einstein's relativity (see for example \cite{Akarsu10, Campo12}). In many such earlier works, the dark energy pressure is considered to be different in different directions which result into directional dark energy equation of states. The source of pressure anisotropy in all those works are not specifically mentioned. In a recent work \cite{SKT15}, we have investigated Bianchi V dark energy models with pressure anisotropy considering only the contribution coming from dark energy in a scale invariant theory proposed by Wesson \cite{Wesson81a, Wesson81b}. The reason behind the formulation of such a theory is that matter in the universe represented by the galaxies appears to be describable in a scale free manner. In other words, the cluster of galaxies and the distribution of galaxies over large distances can be described mathematically through functions which should be free of any fixed space or time scales \cite{Wesson81a}. Another reason behind the proposition of a scale invariant theory of gravity is that, basically it is a gauge theory and gauge theories in particle physics provide better explanation for interaction between particles. Even though several scale invariant theories of gravity have been proposed with varying degrees of success in explaining cosmological observations, Wesson in his work \cite{Wesson81a} has claimed that, his theory is a simple one and may be superior to others. Scale invariant theory of gravity has played important roles in elementary particle physics and cosmology \cite{Dirac74, Canuto77}. Beesham \cite{Bees86},  Mohanty and Mishra\cite{Mohanty03},  Mishra et al. \cite{Mishra04,Mishra12a, Mishra12b, Mishra14}, Barrow \cite{Barrow06},  and Shaposhnikov et al. \cite{Shapo09} have investigated several aspects of this theory. 

With the intention to extend our earlier work \cite{SKT15}, we have constructed some anisotropic dark energy models at the background of an anisotropic and spatially homogeneous Bianchi type $V$ (BV) universe in Wesson's theory. In the present investigation, we have considered two different fluids contributing to the matter field. The first one is the usual bulk viscous cosmic fluid and the second one is due to the dark energy. The two fluids are considered to be non interacting. As in the previous work \cite{SKT15}, pressure anisotropies for dark energy contributions are assumed along different spatial directions. 

The paper is organised as follows: In Section 2, the basic formalism of the theory for an anisotropic $BV$ metric is discussed. The anisotropy in the cosmic fluid is thought to be due to different dark energy pressures along different spatial directions. In Section 3, the skewness parameters are derived from the field equations.  Section-4 contains a small discussion on the anisotropic behaviour of the model. In Section-5, two different cosmological models are constructed by assuming  power law and de Sitter kind of expansion. The dynamical behaviour of the pressure anisotropies are discussed.  We conclude in Section-6. In this paper, gravitational units $8\pi G=c=1$ are used.

\section{Basic Formalism}

In Wesson's theory \cite{Wesson81a}, the field equations are written in a scale-independent way through a conformal   transformation

\begin{equation}
\bar{g}_{ij}=\beta^{2} (x^{k})g_{ij}. \label{eq:1}
\end{equation}

The action principle can be written for such transformation as

\begin{equation}
\delta \int d^4x \sqrt{-g}\left(\beta^2 R+2\mathcal{L}_m\right)=0, \label{eq:1a}
\end{equation}
where  $R=g^{ij}R_{ij}$ is the Ricci scalar,  $R_{ij}$ is the Ricci tensor and $\beta$ is a gauge function. $\mathcal{L}_m$ is the matter Lagrangian density and $g= det (g_{ij})$ is the determinant of the metric tensor. The energy momentum tensor $T_{ij}$ is related to the matter Lagrangian density as 
\begin{equation}
T_{ij}=\beta^{-2}\frac{2}{\sqrt{-g}}\frac{\delta}{\delta g_{ij}}\left[\left(\sqrt{-g} \mathcal{L}_m\right)\right],
\end{equation}
which is a generalisation of 
$\bar{T}_{ij}=\frac{2}{\sqrt{-\bar{g}}}\frac{\delta}{\delta \bar{g}_{ij}}\left[\left(\sqrt{-\bar{g}} \mathcal{L}_m\right)\right]$
in general relativity (GR). A corresponding transformation of the Ricci tensor leads to the Einstein tensor 
\begin{equation}
\bar{G}_{ij}= G_{ij} + 2 \frac{\beta_{;ij}}{\beta} -
4\frac{\beta_{,i}\beta_{,j}}{\beta^2}+(g^{ab}\frac{\beta_{,a}\beta_{,b}}{\beta^2}-2g^{ab}\frac{\beta_{;ij}}{\beta})g_{ij}. \label{eq:2}
\end{equation}

The field equations in presence of an equivalent dimensionless cosmological constant $\Lambda_{0}$ are transformed into

\begin{equation}
G_{ij} + 2 \frac{\beta_{;ij}}{\beta} -
4\frac{\beta_{,i}\beta_{,j}}{\beta^2}+(g^{ab}\frac{\beta_{,a}\beta_{,b}}{\beta^2}-2g^{ab}\frac{\beta_{;ij}}{\beta})g_{ij}+\Lambda_{0}\beta^{2}g_{ij}= -T_{ij}, \label{eq:2a}
\end{equation}
where $G_{ij}= R_{ij}-\frac{1}{2}Rg_{ij}$ is the Einstein tensor. Semicolon denotes covariant differentiation with respect to the metric $g_{ij}$ and comma represents partial differentiation with respect to the coordinates.

We consider Wesson's interval with a Dirac gauge expressed for an anisotropic $BV$ metric as

\begin{equation}
ds^{2}_{W}=\beta^{2}ds^{2}_{E} .\label{eq:4}
\end{equation}
with
\begin{equation}
ds^{2}_{E}=-dt^{2}+A^{2}dx^{2}+e^{2 \alpha x}(B^{2}dy^{2}+C^{2}dz^{2}).\label{eq:5}
\end{equation}
Here, $A,B,C$ are functions of $t$ only.

The average scale factor $a$ and volume scale factor $V$ for the BV model are $a=(ABC)^\frac{1}{3}$ and $V=a^3=ABC$. The mean Hubble's parameter is $H=\frac{\dot{a}}{a}=\frac{1}{3}(H_x+H_y+H_z)$, where $H_x=\frac{\dot{A}}{A},H_y=\frac{\dot{B}}{B}, H_z=\frac{\dot{C}}{C}$ are the respective directional Hubble parameters. 

The energy momentum tensor for an environment with two non interacting fluids can be written as

\begin{equation}
T_{ij}=T_{ij}^{(vis)}+T_{ij}^{(de)}\label{eq:6}
\end{equation}
where $T_{ij}^{(vis)}$ and $T_{ij}^{(de)}$ respectively denote the contribution to the energy momentum tensor from bulk viscous cosmic fluid and dark energy respectively. It may be mentioned here that bulk viscous fluid may play the role of dark energy and can explain the cosmic speed up phenomena \cite{Brevik17, SKT10}. However, in the present work, we are interested to investigate the anisotropic nature and role of dark energy through the assumption of anisotropic pressure along different directions in presence of an isotropic bulk viscous fluid.

For a viscous cosmic fluid, we have
\begin{equation}
T_{ij}^{(vis)}=(\rho+\bar{p})u_iu_j+\bar{p}g_{ij}.\label{eq:7}
\end{equation}
where $\bar{p}=p-\zeta u_{;i}^i$, so that $T_{ij}^{(vis)}=diag[-\rho, p, p, p]$.  In comoving coordinates, $u^i = \delta^i_0$ is the four velocity vector and $\zeta$ is the bulk viscosity coefficient. $\zeta u_{;i}^i = 3\zeta H$ is the bulk viscous pressure. 

We consider the dark energy contribution as

\begin{eqnarray}
T_{ij}^{(de)} &=& diag[-\rho_d, p_{xd}, p_{yd},p_{zd}]\nonumber \\
         & =& diag[-1, \omega_{dx}, \omega_{dy},\omega_{dz}]\rho_d \nonumber \\
         & =& diag[-1, (\omega_d+\delta),(\omega_d+\gamma), (\omega_d+\eta)]\rho_d,\label{eq:8}
\end{eqnarray}
where $\omega_d$ is the dark energy equation of state parameter (EoS) and $\rho_d$ is the dark energy density. $\delta$, $\gamma$, and $\eta$ are respectively the departures from $\omega_d$ along $x$, $y$ and $z$ axes.

The field equations become

\begin{equation}
\frac{\ddot{B}}{B}+\frac{\ddot{C}}{C}+\frac{\dot{B}\dot{C}}{BC}-\frac{\alpha^2}{A^2}+2\frac{\dot{\beta}}{\beta}\biggl(\frac{\dot{B}}{B}+\frac{\dot{C}}{C}\biggr)
+2\frac{\ddot{\beta}}{\beta}-\frac{\dot{\beta}^2}{\beta^2}+\Lambda_0\beta^2= -p+3\zeta H-(\omega_d+\delta)\rho_d.\label{eq:9}
\end{equation}

\begin{equation}
\frac{\ddot{A}}{A}+\frac{\ddot{C}}{C}+\frac{\dot{A}\dot{C}}{AC}-\frac{\alpha^2}{A^2}+2\frac{\dot{\beta}}{\beta}\biggl(\frac{\dot{A}}{A}+\frac{\dot{C}}{C}\biggr)
+2\frac{\ddot{\beta}}{\beta}-\frac{\dot{\beta}^2}{\beta^2}+\Lambda_0\beta^2=-p+3\zeta H -(\omega_d+\gamma)\rho_d .\label{eq:10}
\end{equation}

\begin{equation}
\frac{\ddot{A}}{A}+\frac{\ddot{B}}{B}+\frac{\dot{A}\dot{B}}{AB}-\frac{\alpha^2}{A^2}+2\frac{\dot{\beta}}{\beta}\biggl(\frac{\dot{A}}{A}+\frac{\dot{B}}{B}\biggr)+ 2\frac{\ddot{\beta}}{\beta}-\frac{\dot{\beta}}{\beta^2}+\Lambda_0\beta^2= -p+3\zeta H-(\omega_d+\eta)\rho_d .\label{eq:11}
\end{equation}

\begin{equation}
\frac{\dot{A}\dot{B}}{AB}+\frac{\dot{B}\dot{C}}{BC}+\frac{\dot{C}\dot{A}}{CA}-3\frac{\alpha^2}{A^2}+
2\frac{\dot{\beta}}{\beta}\biggl(\frac{\dot{A}}{A}+\frac{\dot{B}}{B}+\frac{\dot{C}}{C}\biggr)+3\frac{\dot{\beta}^2}{\beta^2}+\Lambda_0\beta^2=\rho+\rho_d.\label{eq:12}
\end{equation}

\begin{equation}
2\frac{\dot{A}}{A}-\frac{\dot{B}}{B}-\frac{\dot{C}}{C}=0 .\label{eq:13}
\end{equation}

The overhead dot denotes ordinary time derivatives. One should note that, in the absence of bulk viscous cosmic fluid, the above equations reduce to the corresponding equations in Ref. \cite{SKT15}. 

The scalar expansion for the model is $\theta=3H=\frac{\dot{V}}{V}=\frac{\dot{A}}{A}+\frac{\dot{B}}{B}+\frac{\dot{C}}{C}$ and shear scalar is  $\sigma^2=\frac{1}{2}\sigma_{ij}\sigma^{ij}=\frac{1}{2}\biggl(\Sigma H_i^2-\frac{1}{3}\theta^2\biggr)$. $H_i$ with $i=x,y,z$ are the respective directional Hubble parameters. 


For a spatially homogeneous metric, a linear relationship among the scalar expansion and shear scalar is usually considered  \cite{Collins80, BMSKT15, SKT09} that leads to an anisotropic relation $B=C^m$ where $m\neq 1$ is an arbitrary positive constant. Integration of eqn. $\eqref{eq:13}$ yields $A=(BC)^\frac{1}{2}$ and consequently the directional scale factors can be expressed as $A=a$, $B=C^m=a^{\frac{2m}{m+1}}$. We can have $H_x=H$, $H_y=\left(\frac{2m}{m+1}\right)H$ and $H_z=\left(\frac{2}{m+1}\right)H$. It may be noted here that, along the $x-$axis, the directional Hubble parameter is the same as that of the mean Hubble parameter which implies that along  $x-$axis, the deviation of the anisotropic dark energy equation of state from mean equation of state is minimum or zero i.e $\delta$ should vanish.


\section{Skewness parameters and Pressure Anisotropy}
The conservation equation $T^{ij}_{;j}=0 $ yields

\begin{equation}
\dot{\rho}+3(\bar{p}+\rho)H+\dot{\rho_d}+3\rho_d(\omega_d+1)H+\rho_d(\delta H_x+\gamma H_y+\eta H_z)=0.\label{e21}
\end{equation}
We consider here a non interacting dark energy in presence of a viscous cosmic fluid which leads to the splitting of the above equation into two parts: one corresponding to the usual viscous fluid with isotropic pressure and the other for dark energy i.e 
\begin{equation}
\dot{\rho}+3(\bar{p}+\rho)H=0 \label{eq:17}
\end{equation}
and 

\begin{equation}
\dot{\rho_d}+3\rho_d(\omega_d+1)H+\rho_d(\delta H_x+\gamma H_y+\eta H_z)=0. \label{eq:18}
\end{equation}

The conservation equation \eqref{eq:18} can  further be split into the isotropic part and the anisotropic part.

\begin{equation}\label{eq:19}
\dot{\rho_d}+3\rho_d(\omega_d+1)H=0,
\end{equation}
and
\begin{equation}\label{eq:20}
\rho_d(\delta H_x+\gamma H_y+\eta H_z)=0.
\end{equation}
The second equation deals with the deviations from the total dark energy pressure along different spatial directions. According to eqn. \eqref{eq:19}, the behaviour of $\rho_d$ depends the deviation free part of dark energy EoS parameter $\omega$: $\rho_d=\rho_{d0} a^{-3(\omega_d+1)}$. At the present epoch, $\rho_{d}$ reduces to be $\rho_{d0}$.

Algebraic manipulation of the equations \eqref{eq:9}, \eqref{eq:10} and \eqref{eq:11} yields
\begin{equation}
(m-1)F(H,\beta)=(\gamma -\eta)\rho_d\label{e24}
\end{equation}
and
\begin{equation}
\left(\frac{m-1}{2}\right)F(H,\beta)=(\gamma -\delta)\rho_d,\label{e33}
\end{equation}
where  $F(H,\beta)=\frac{2}{m+1}\left(\dot{H}+3H^2+2H\frac{\dot{\beta}}{\beta}\right)$ and $\gamma =2\delta -\eta$ .

The skewness parameters are obtained from equations \eqref{eq:20}, \eqref{e24} and \eqref{e33} as

\begin{eqnarray}
\delta &=& -\left(\frac{m-1}{3\rho_d}\right)\chi(m)F(H,\beta),\label{e35}\\
\gamma &=& \left(\frac{5+m}{6\rho_d}\right)\chi(m)F(H,\beta),\label{e36}\\
\eta &=& -\left(\frac{5m+1}{6\rho_d}\right)\chi(m)F(H,\beta),\label{e37}
\end{eqnarray}
where, $\chi(m)=\frac{m-1}{m+1}$ which is a measure of departure from isotropic nature. $\chi(m)$ vanishes for isotropic model and consequently the pressures along all spatial directions become equal.

The dark energy density $\rho_d$ and  the EoS parameter $\omega_d$ are obtained as
\begin{eqnarray}
\rho_d & = &2\left(\frac{m^2+4m+1}{(m+1)^2}\right)H^2-\frac{3\alpha^2}{a^2}+6H\frac{\dot{\beta}}{\beta}\nonumber\\
       & + &3\left(\frac{\dot{\beta}}{\beta}\right)^2+\Lambda_0\beta^2-\rho,\label{e38}\\
\omega_d\rho_d & = &-\bar{p}-\frac{2}{3}\left(\frac{m^2+4m+1}{m+1}\right) \left[F(H,\beta)-\frac{3H^2}{m+1}\right] + \frac{\alpha^2}{a^2}\nonumber\\
               & - & 2\frac{\ddot{\beta}}{\beta}+ \left(\frac{\dot{\beta}}{\beta}\right)^2-\Lambda_0\beta^2.\label{e39}
\end{eqnarray}
The behaviour of the skewness parameters are decided by the functional $F(H,\beta)$ and  the dark energy density $\rho_d$. In fact, the skewness parameters depend on the ratio $\frac{F(H,\beta)}{\rho_d}$. $F(H,\beta)$ depends on $m$, $H$ and $\beta$. The behaviours of dark energy density and dark energy EoS parameter are decided by the anisotropic nature of the model through the exponent $m$, the time dependence of mean Hubble parameter $H$, the gauge function $\beta$ and the rest energy density $\rho$. It is interesting to note that, the dark energy density and the dark energy EoS parameter depend on the usual cosmic fluid. In the absence of usual cosmic fluid, the above equations \eqref{e38} and \eqref{e39} reduce to the earlier expressions of Ref. \cite{SKT15}.

The scale invariant theory as proposed by Wesson reduces to the usual General Relativity theory for a constant gauge function $\beta$ and a vanishing cosmological constant $\Lambda_0$. Assuming a constant gauge function and setting $\Lambda_0=0$ in the eqs.\eqref{e38} and \eqref{e39}, we obtain,

\begin{eqnarray}
\rho_d & = &2\left(\frac{m^2+4m+1}{(m+1)^2}\right)H^2-\frac{3\alpha^2}{a^2}-\rho,\label{e71}\\
\omega_d\rho_d & = &-\bar{p}-\frac{2}{3}\left(\frac{m^2+4m+1}{m+1}\right) \left[F(H)-\frac{3H^2}{m+1}\right] + \frac{\alpha^2}{a^2} .\label{e72}
\end{eqnarray}
Here, $F(H)=\frac{2}{m+1}\left(\dot{H}+3H^2\right)$. In the absence of a bulk viscous cosmic fluid, the above equations reduce to the expressions as obtained in Ref.\cite{BMSKT15}. For a barotropic bulk viscous pressure $\bar{p}=\varepsilon \rho$ \cite{SKT09,SKT10}, the rest energy density of universe can be expressed as 

\begin{equation}
\rho=\rho_0 e^{-3(1+\varepsilon)\int H dt},\label{eq27}
\end{equation}
where $\rho_0$ is the corresponding value at present epoch. Bulk viscosity has already been recognised as a dissipative phenomena which plays a crucial role in getting accelerated phase of expansion. It is worth to mention here that $\varepsilon$ can either be positive and negative \cite{SKT09,SKT10}. DE density and DE EoS can be calculated from the rest energy density $\rho$ which we can get from \eqref{eq27} for a given value of $\epsilon$ and given nature of $H$.

\section{Anisotropic nature of the model}
Standard cosmological model is based upon the assumption of cosmic isotropy and homogeneity. Breaking of this isotropy leads to Bianchi type models. We have considered such a Bianchi cosmology with spatial homogeneous sections and anisotropy in directional expansion rates. The anisotropic nature of the considered model is assessed through  the calculation of the average anisotropy parameter(AP) $\mathcal{A}$ given by
\begin{equation}
\mathcal{A}=\frac{1}{3}\Sigma \left(\frac{\Delta H_i}{H}\right)^2. \label{eq:14}
\end{equation}
Here $\Delta H_i=H_i-H$ with $i=x,y,z$.

The AP is obtained as \cite{SKT15}
\begin{equation}
\mathcal{A}=\frac{2}{3}\left(\frac{m-1}{m+1}\right)^2.\label{eq:15}
\end{equation}
The average anisotropic parameter becomes a constant quantity. In principle, one can get a time dependent anisotropic parameter with different assumption. In order to get a first hand information about the pressure anisotropies along different spatial directions, we have adopted such a simplified approach. The anisotropic parameter depends only on the exponent $m$ and vanishes for an isotropic case with $m=1$. One may note that, the universe is observed to be mostly flat and isotropic. The observations regarding the spatial anisotropy of the universe should imply a sort of small perturbation to the usual isotropic nature. Accordingly, we have considered the anisotropic exponent to be $m=1.0001633$ which corresponds to an anisotropic parameter $\mathcal{A}= 4.4439\times 10^{-9}$ \cite{SKT15, BMSKT15}. This estimate of anisotropy is close to the value  obtained from WMAP data \cite{Campa11}.

The anisotropic nature of model can also be  quantified through the estimation of $\frac{\sigma}{H}$ at the present epoch. For the BV model, we obtain this quantity as 

\begin{equation}
\frac{\sigma}{H}= \left[\frac{5m^2+2m+5}{2(m+1)^2}-1.5\right]^{\frac{1}{2}}.\label{eq:15a}
\end{equation}
In some recent works, limits to this quantity have been obtained from different observational data. From an analysis of COBE data, Bunn et al. have obtained a limit of $\left(\frac{\sigma}{H}\right)_0 < 3 \times 10^{-9}$ \cite{Bunn96}. However, from the data of cosmic microwave  background temperature and polarisation data from Planck, Saadeh et al. put a tighter constraint of  $\left(\frac{\sigma}{H}\right)_0 < 4.7 \times 10^{-11}$  with the conclusion that anisotropic expansion of the universe is strongly disfavoured \cite{Saadeh16}. The parameter $m$ in our model can be constrained according to these bounds on the anisotropy in expansion rates to a range very close to 1. However, we use the previous constraint on $m$ i.e $m=1.0001633$ so that $\left(\frac{\sigma}{H}\right)_0$ becomes $8.164 \times 10^{-5}$. 

\section{Some Cosmological Models}

In the description of the late time cosmic dynamics, deceleration parameter (DP) $q=-\frac{a\ddot{a}}{\dot{a}^2}$ plays an important role. A positive $q$ implies a decelerating universe and a negative $q$ describes an accelerating universe. In recent times, models with late time cosmic speed up have gained much importance. In the present study, we focus on the late time cosmic dynamics when DP is believed to be a constant or be varying slowly with cosmic time. Different analyses of the observational data from type Ia Supernovae constrain the deceleration parameter in the present epoch to be  $q=-0.81\pm 0.14$ \cite{Rapetti07} or $q=-0.53^{+0.17}_{-0.13}$ \cite{Giostri12}. Keeping in view of the cosmic dynamics, one can choose the behaviour of the scale factor with all plausible assumptions. Common choices for the scale factor are the de Sitter solution and the power law expansion. In de Sitter solution, the scale factor increases exponentially with cosmic time whereas in the power law expansion, the scale factor is chosen to vary as certain power of cosmic time. Power law or an exponential behaviour of the scale factor simulates a constant DP. 

\subsection{Power law expansion}
In this model, we choose to have power law functional form for the scale factor in the form $a=t^{(\frac{m+1}{2})n}$, where the exponent $n$ is an arbitrary positive constant. If we define a constant $k=(\frac{m+1}{2})n$, the scale factor becomes $a=t^k$. The volume scale factor will now behave like $V=t^{3k}$ and the Hubble parameter is $H=\frac{k}{t}$. In terms of redshift $z$, the mean Hubble rate is $H(z)=k(1+z)^{\frac{1}{k}}$. Here we take $1+z=\frac{1}{a}$, where the scale factor in the present epoch is $1$. $k$ is related to the value of Hubble parameter in the present epoch. It is obvious that, for $k>1$, the model will be an accelerating one. The directional Hubble rates are $H_x=\left(\frac{m+1}{2}\right)\frac{n}{t}$, $H_y=\frac{mn}{t}$ and $H_z=\frac{n}{t}$. DP becomes $q=-1+\frac{2}{n(m+1)}$ and is negative for $n > \frac{2}{m+1}$. In order to get an accelerating model with this power law scale factor, the exponent $n$ should always be greater than one i.e. $n>1$ if $m<1$ otherwise  it has to be decided from $n > \frac{2}{m+1}$. The universe is in general isotropic but recent observations hint towards an asymmetric spatial expansion. However, anisotropy in spatial expansion must be very small and therefore the exponent $m$ must be very close to $1$. In fact, in the present model, from the analysis of anisotropy as predicted from WMAP data, $m=1.0001633$. This further restricts the value of the parameter $n$ to be very close to 1.

The gauge function $\beta$ may be chosen suitably for getting a viable model. It may depend either on cosmic time or the scale factor. We consider a gauge function which behaves reciprocally with the cosmic time, so that at late phase, the function will have a little contribution. It is necessary to mention here that, such a choice retains the dimensional consistency of the gauge function $\beta$ as has been pointed out by Wesson \cite{Wesson81a}. 

The energy density contribution coming from the usual cosmic fluid for the power law model reduce to 

\begin{equation}
\rho=\frac{\rho_0}{t^{\frac{3}{2}(1+\varepsilon)(m+1)n}}. \label{e69}
\end{equation}
Now, with the choice of $\beta=\frac{1}{t}$, the functional $F(H,\beta)$ becomes a function of $t$ i.e. $F(t)$.  The functional $F(t)$ and the dark energy density $\rho_d$ are obtained as

\begin{eqnarray}
F(t) &=& \left[\frac{n^2(m+1)}{2}-n\right]\frac{3}{t^2},\label{e40}\\
\rho_d &=& \left[\frac{(m^2+4m+1)n^2}{2}-3n(m+1)+(3+\Lambda_0)\right]\frac{1}{t^2}\\
      &-&\frac{3\alpha^2}{t^{n(m+1)}}-\frac{\rho_0}{t^{\frac{3}{2}(1+\varepsilon)(m+1)n}} .\label{e41}
\end{eqnarray}


The dark energy EoS parameter $\omega_d$ becomes 

\begin{eqnarray}
\omega_d \rho_d & = & -\left[\frac{m^2+4m+1}{m+1}\left(\frac{(m+1)n^2}{2}-2n\right)+\left(3+\Lambda_0\right)\right]\frac{1}{t^2}\\ \nonumber
                & + &\frac{\alpha^2}{t^{n(m+1)}}-\frac{\varepsilon\rho_0}{t^{\frac{3}{2}(1+\varepsilon)(m+1)n}} .\label{45}
\end{eqnarray}

The functional $F(t)$ has the same form as that of Ref. \cite{SKT15} and is not affected by the presence of a usual bulk viscous cosmic fluid. As usual, it behaves as $t^{-2}$. The signature of $F(t)$ is positive for $n> \frac{2}{m+1}$ and negative if $n< \frac{2}{m+1}$.  Hence, it decreases with time for $n> \frac{2}{m+1}$ and increases for $n< \frac{2}{m+1}$.  The functional $F(t)$ vanishes for the critical relationship $n(m+1)=2$ and consequently all the skewness parameters vanish implying an isotropic dark energy pressure in all spatial directions. We are interested in the late time cosmic speed up and therefore we take $n> \frac{2}{m+1}$ for the power law expansion which makes the functional $F(t)$ to be positive. $\rho_d$ is found to decrease with time. Three factors decides the rate of decrement in $\rho_d$: $t^{-2}$, $t^{-n(m+1)}$ and $t^{-\frac{3}{2}(1+\varepsilon)(m+1)n}$ in respective terms. The role of bulk viscous cosmic fluid comes through the third term. One may note that, if $\varepsilon=-1$, even though the contribution coming from the usual cosmic fluid does not vanish, it does not contribute to the time variation of $\rho_d$. For $\varepsilon=-\frac{1}{3}$, the time variation of second and third terms can be clubbed together. For $n(m+1)=2$ and $\varepsilon=-\frac{1}{3}$, the dark energy density $\rho_d$ behaves as $t^{-2}$ and hence the ratio $\frac{F(t)}{\rho_d}$ becomes independent of time. Consequently, for this choice the skewness parameters becomes a constant quantity appearing as  simple time independent deviations from usual isotropic pressure. This situation can also arise for $n=\frac{2}{m+1}$ and $\varepsilon=-1$. If $n(m+1)>2$ and $\varepsilon>-\frac{1}{3}$, the magnitude of decrement in the dark energy density becomes more rapid compared to the decrement in $F(t)$. Accordingly, the ratio $\frac{F(t)}{\rho_d}$ for the particular case will increase with cosmic time. However, for large value of $n$, the ratio $\frac{F(t)}{\rho_d}$ behaves as unity. Therefore, for large value of $n$, the skewness parameters depend only on the parameter $m$ and become independent of cosmic time.

\begin{figure}[h!]
\begin{center}
\includegraphics[width=1\textwidth]{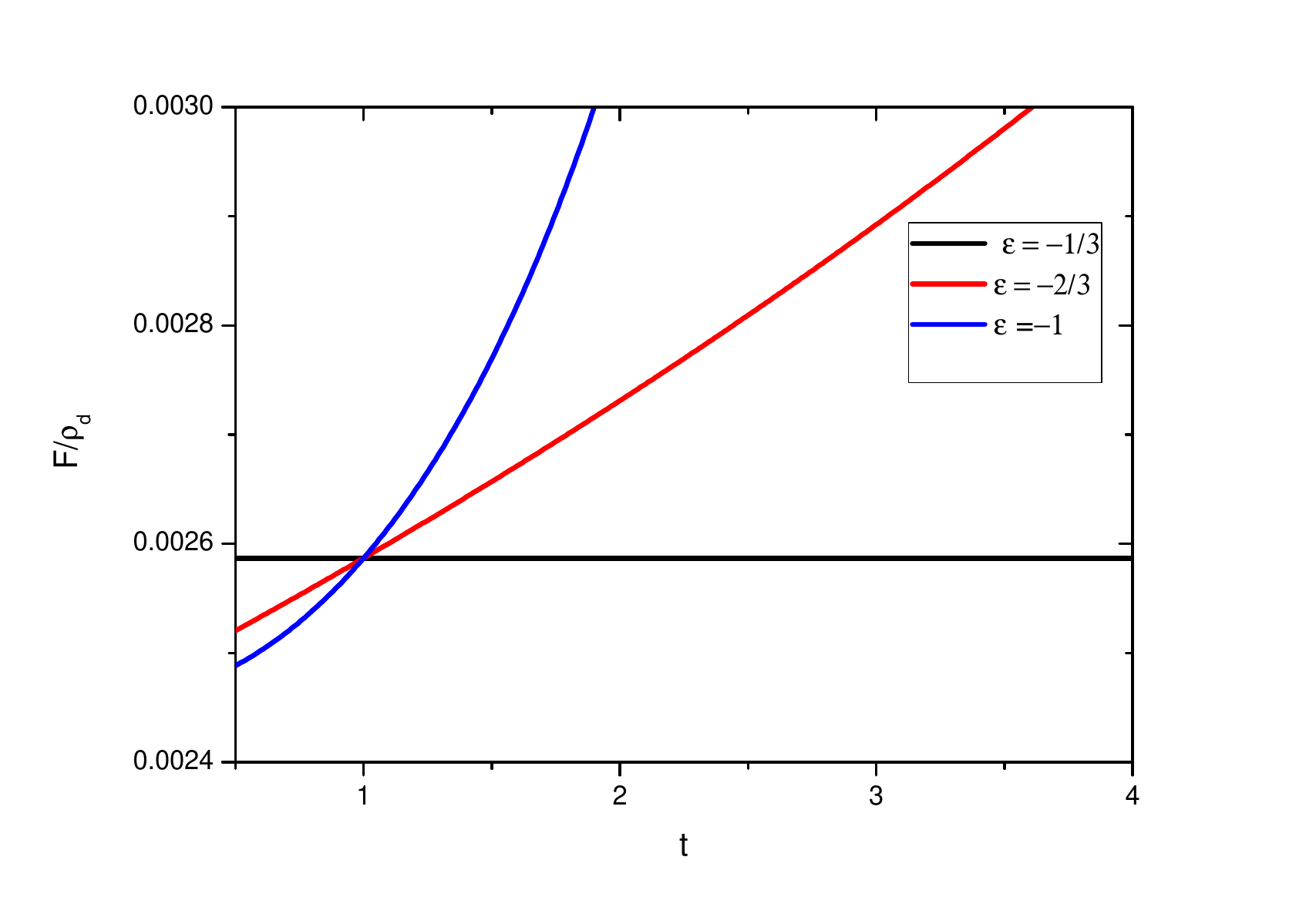}
\caption{Evolution of the functional $\frac{F(t)}{\rho_d}$  with cosmic time for the power law model.}
\end{center}
\end{figure}

\begin{figure}[h!]
\begin{center}
\includegraphics[width=1\textwidth]{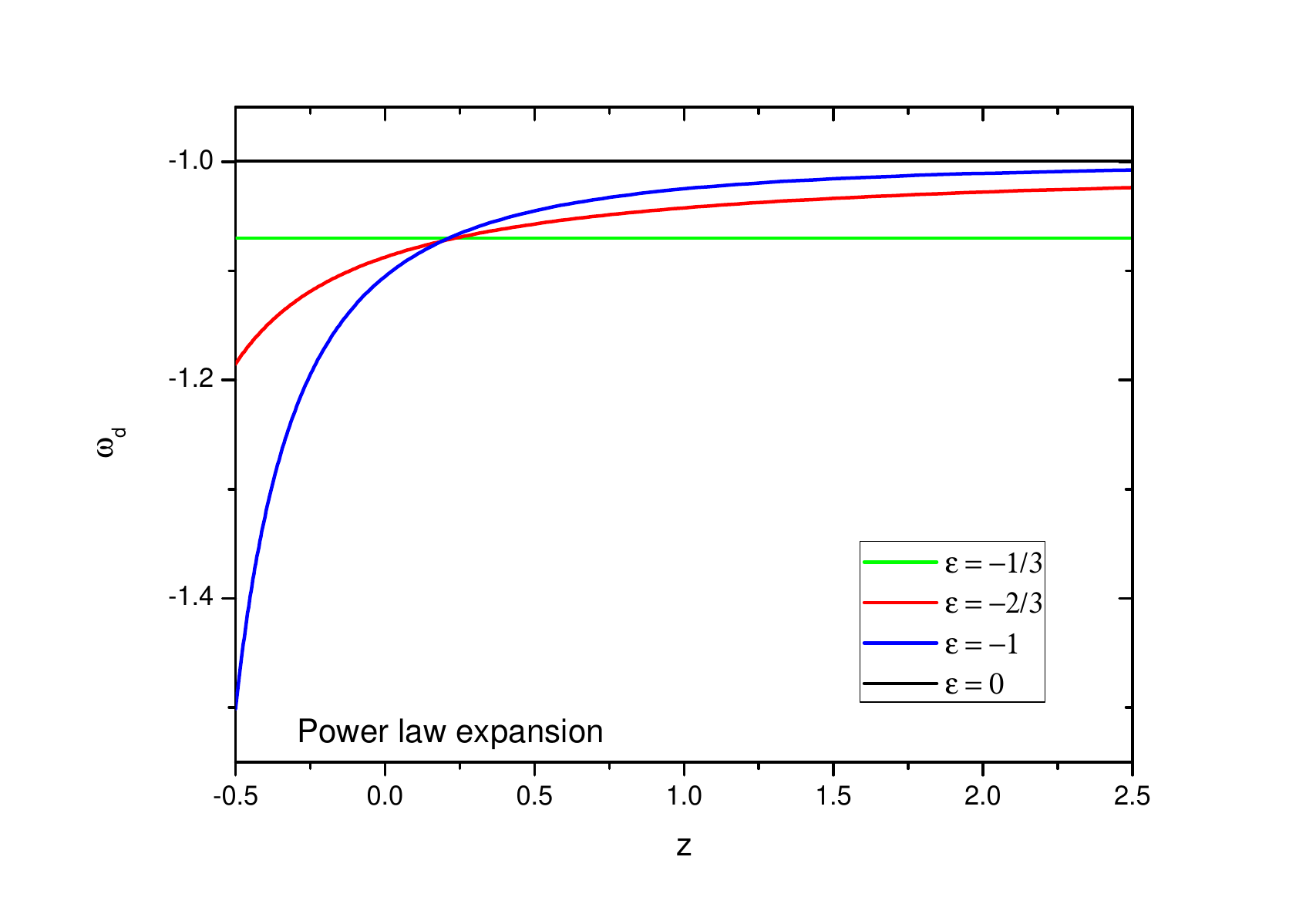}
\caption{Dynamics of the dark energy equation of state parameter  $\omega_d$ for power law expansion. Four representative values of the parameter $\varepsilon$ are considered. $\varepsilon =0$ represents a universe with no usual bulk viscous matter but of only dark fluid.}
\end{center}
\end{figure}

\begin{figure}[h!]
\begin{center}
\includegraphics[width=1\textwidth]{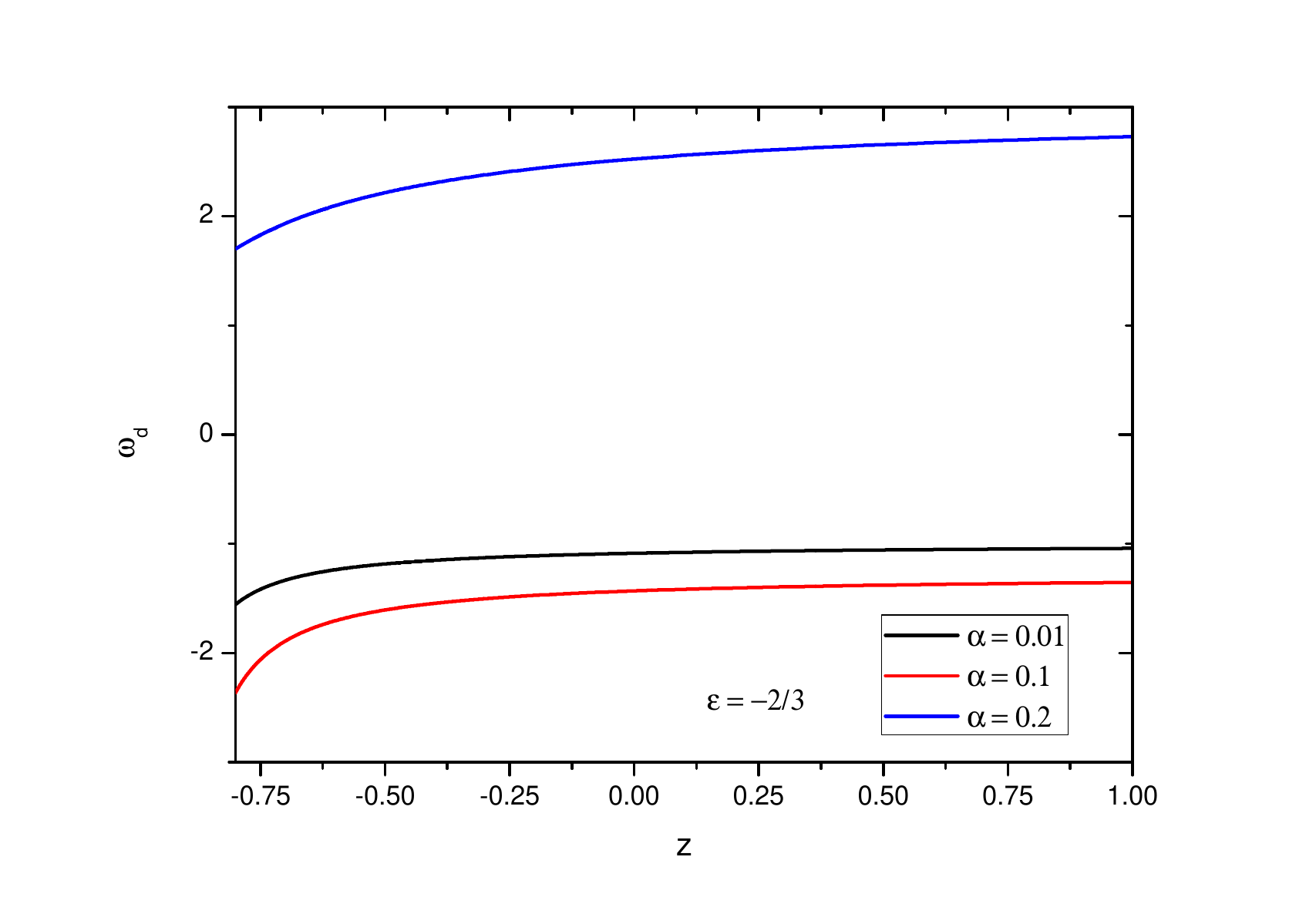}
\caption{Variation of $\omega_d$ with the parameter $\alpha$ of the anisotropic metric BV for power law model of expansion.}
\end{center}
\end{figure}

In Figure 1 , we have plotted the ratio $\frac{F(t)}{\rho_d}$ for three representative values of $\varepsilon$, namely $ \varepsilon=-\frac{1}{3}, -\frac{2}{3}$ and $-1$. As discussed above, except for the specific choice $ \varepsilon=-\frac{1}{3}$, the ratio $\frac{F(t)}{\rho_d}$ increases with cosmic time. The slope of the curves increases with the decrease in the value of $\varepsilon$. In otherwords, the curve for $\varepsilon=-1$ becomes more stiff than the curve with $\varepsilon=-\frac{2}{3}$. $\frac{F(t)}{\rho_d}$ is almost time independent for $ \varepsilon=-\frac{1}{3}$. It can be observed from the figure that, for all considered values of $\varepsilon$, $\frac{F(t)}{\rho_d}$ becomes equal at the present cosmic time $t=1$. Since the behaviour of the skewness parameters depends mostly on the behaviour of the ratio $\frac{F(t)}{\rho_d}$, we can easily assess them from Fig.1.

In Figure-2, the evolution of dark energy equation of state $\omega_d$ is shown for four different values of $\varepsilon$, namely $ \varepsilon=0,-\frac{1}{3}, -\frac{2}{3}$ and $-1$. The case $\varepsilon=0$ corresponds to the cosmic fluid with no usual matter but of only dark fluid. For this particular case, the dark fluid behaves like a cosmological constant with $\omega_d=-1$. $\omega_d$  remains in the phantom region for all the cases considered here. As expected, for $\varepsilon=-\frac{1}{3}$, $\omega_d$ is independent of time but it remains in the phantom region below the phantom divide. In the remaining two cases, $\omega_d$ decreases with cosmic time. At early times, the model behaves like a cosmological constant for these cases of $\varepsilon$. At late times, $\omega_d$ decreases to acquire larger negative value. However, the present model with two non interacting fluids favour phantom field to dominate the cosmic dynamics. One can note from the figure that, $\omega_d$ decreases more rapidly for lower values of $\varepsilon$ after $\varepsilon=-\frac{1}{3}$. One can note that at certain redshift around $z=0.25$, $\omega_d=-1.07$ for all the three cases of $\varepsilon$ except $\varepsilon=0$.

$\omega_d$ is sensitive to the choice of the parameters $\alpha$ and $n$.  The variation of $\omega_d$ with respect to $\alpha$ is plotted in Figure-3 and with respect to $n$ is shown in Figure-4. For all the choices of $\alpha$, the general trend in $\omega_d$ is the same even though, it remains in the positive domain for some higher values of $\alpha$. $\omega_d$ shows an increasing trend with the increase in $n$. For some higher values of $n$, $\omega_d$ becomes positive all through the cosmic evolution.

\begin{figure}[h!]
\begin{center}
\includegraphics[width=1\textwidth]{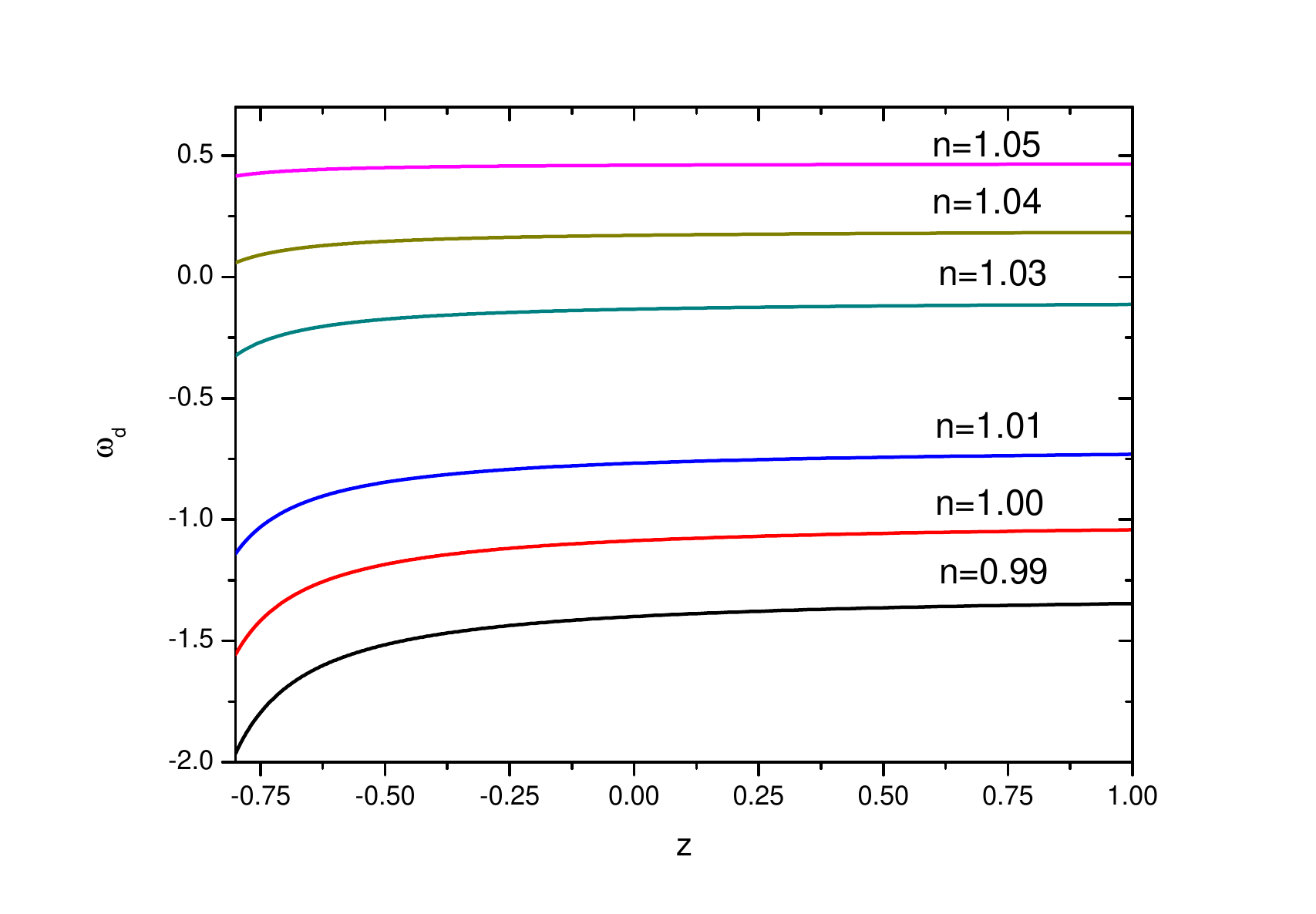}
\caption{Variation of $\omega_d$ with the parameter $n$ of the exponent in the power law model of expansion.}
\end{center}
\end{figure}

\begin{figure}[h!]
\begin{center}
\includegraphics[width=1\textwidth]{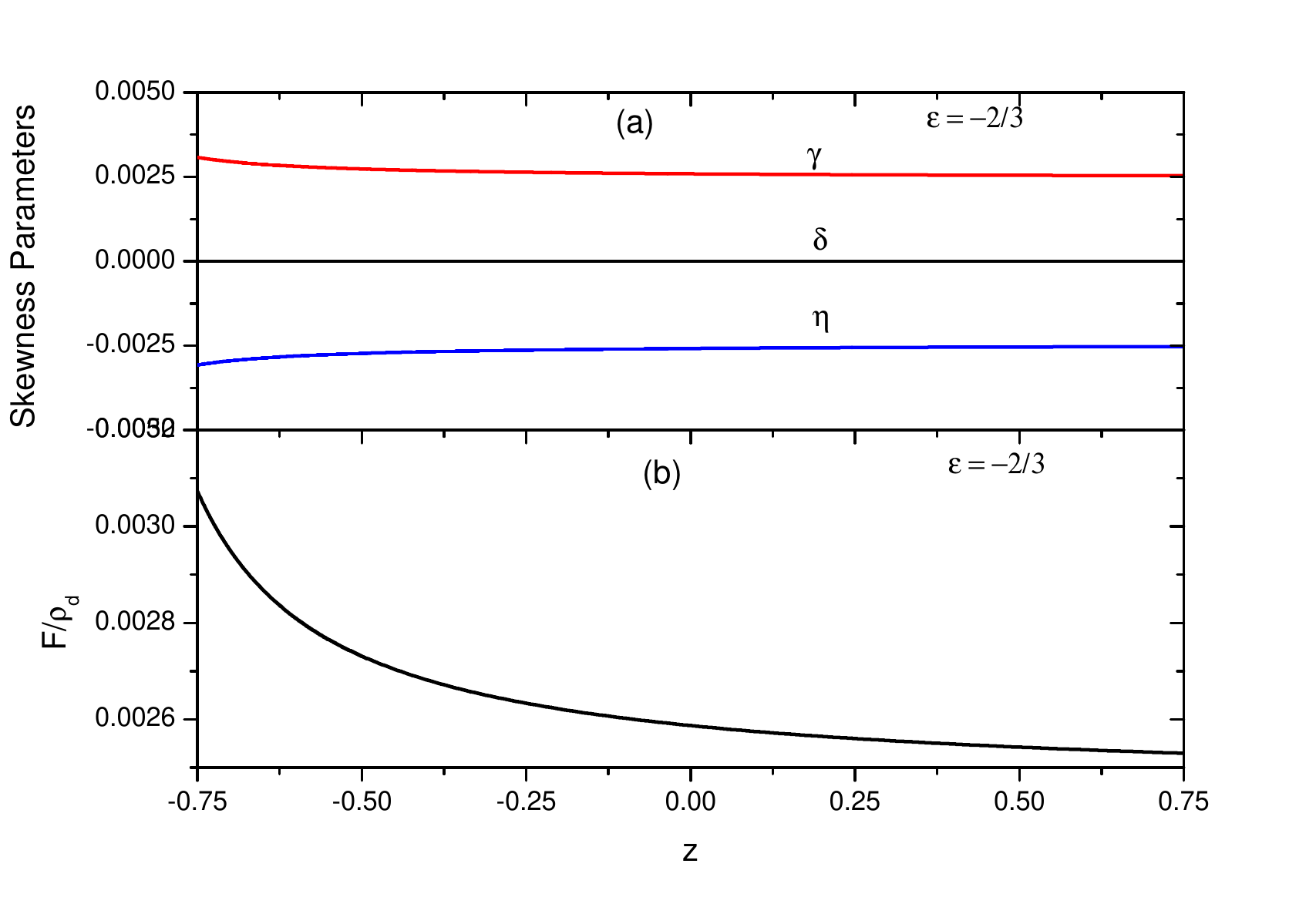}
\caption{(a)(upper panel) Dynamical evolution of skewness parameters along three spatial directions with respect to redshift for the power law model of expansion only  for the case  $\varepsilon =-2/3$. (b)  (lower panel) Since the behaviour of  $\frac{F(t)}{\rho_d}$ controls the nature of the pressure anisotropies in dark energy fluid, its evolution  as function of redshift is shown for $\varepsilon =-2/3$ for reference.}
\end{center}
\end{figure}

The skewness parameters normalised to the functional $\chi(m)$ are shown in Figure-5 as function of redshift. Since the behaviour of the functional $\frac{F}{\rho_d}$ is the same for all the cases of $\varepsilon$, we have only plotted the skewness parameters for a representative $\varepsilon =-\frac{2}{3}$. The skewness parameters closely depend on the behaviour of $\frac{F}{\rho_d}$. Therefore, we have shown $\frac{F}{\rho_d}$ for $\varepsilon =-\frac{2}{3}$ as a function of $z$ alongside of the skewness parameters for a quick reference.  The behaviour of $\frac{F}{\rho_d}$ is clearly reflected in the behaviour of skewness parameters. As expected, the skewness parameter along x-axis , $\delta$, vanishes implying that along x-axis, the cosmic fluid has a pressure equal to the total pressure.  The pressure anisotropies maintain almost a constant value from some early  times to considerably late phase of cosmic evolution. At large cosmic time, the magnitudes of pressure anisotropies tend to increase. It is certain that, the behaviour of $\eta$ is just opposite to that of $\gamma$. $\gamma$ is positive whereas $\eta$ is negative all through the cosmic evolution. Interestingly $\delta,\gamma, \eta$ and $\omega_d$ become independent of time for $n=\frac{2}{m+1}$. In the absence of scale invariance, $\omega_d$ remains completely in the phantom region with much larger negative values than the scale invariant model. In the absence of the cosmological constant, $\omega_d$ is found to remain in the  positive domain throughout the cosmic expansion history.
\subsection{de Sitter expansion}

In de Sitter model, the scale factor $a$ is taken as $a=e^{\frac{(m+1)\xi t}{2}}$ and the volume scale factor behaves as $V=e^{\frac{3(m+1)\xi t}{2}}$. $\xi=\frac{2H}{m+1}$ is a positive constant. In this model, $H$ remains constant through out the cosmic evolution and consequently, DP becomes $q=-1$. The directional Hubble rates along different spatial directions are given by $H_x=\left(\frac{m+1}{2}\right)\xi, H_y=m\xi$ and $H_z=\xi$. These directional Hubble rates are also constant quantities.

The energy density contribution coming from the usual viscous cosmic fluid for the de Sitter model reduce to 

\begin{equation}
\rho=\frac{\rho_0}{e^{\frac{3}{2}(1+\varepsilon)(m+1)\xi t}}. \label{e70}
\end{equation}

The energy density depends on the value of $\varepsilon$.  $\rho$ decreases with the increase in 
$\varepsilon$ and the other way around for a decrement in $\varepsilon$. For the particular choice  $\varepsilon=-1$, $\rho$ becomes independent of time and assumes a constant value $\rho_0$ throughout the cosmic evolution. As in the previous model, we consider the same gauge function i.e. $\beta=\frac{1}{t}$. The functional $F(t)$ and $\rho_d$ for this model can be obtained as

\begin{eqnarray}
 F(t) &=& \frac{3\xi^2(m+1)}{2}-\frac{2\xi}{t},\label{e51}\\
 \rho_d  &=& \frac{(m^2+4m+1)\xi^2}{2}-\frac{3\xi(m+1)}{t}+\frac{(3+\Lambda_0)}{t^2}\\\nonumber
        &-& \frac{3\alpha^2}{e^{\xi(m+1)t}}-\frac{\rho_0}{e^{\frac{3}{2}(1+\varepsilon)(m+1)\xi t}}.\label{e52}
\end{eqnarray}

The dark energy EoS parameter is expressed as

\begin{eqnarray}
\omega_d \rho_d & = & -\left(\frac{m^2+4m+1}{m+1}\right)\left[\frac{(m+1)\xi^2}{2}-\frac{4\xi}{3t}\right]-\frac{\left(3+\Lambda_0\right)}{t^2}\\ \nonumber
                & + &\frac{\alpha^2}{e^{\xi(m+1)t}}-\frac{\varepsilon \rho_0}{e^{\frac{3}{2}(1+\varepsilon)(m+1)\xi t}} .\label{e56}
\end{eqnarray}


In the de Sitter model, $F(t)$ evolves from an early  negative value to a positive value. The growth rate of $F(t)$ is governed by $t^{-1}$. Bulk viscosity does not affect this functional and hence for all the choices of $\varepsilon$, its behaviour remains the same. The dark energy density $\rho_d$ decreases with cosmic time. Four different factors $t^{-1}$, $t^{-2}$, $t^{-n(m+1)}$ and $t^{-\frac{3}{2}(1+\varepsilon)(m+1)n}$ decide the decrement in $\rho_d$. The role of bulk viscous cosmic fluid comes through the fourth term. The contribution from the bulk viscous cosmic fluid becomes time independent for $\varepsilon=-1$. The directional pressure anisotropies depend on the behaviour of $F(t)$ and $\rho_d$ and more specifically on $\frac{F}{\rho_d}$. One should note that, $F(t)$ is not affected by the presence of the bulk viscous cosmic fluid but the dark energy density $\rho_d$ depends on $\varepsilon$ and hence the ratio $\frac{F}{\rho_d}$ depends on $\varepsilon$. We have shown the time variation of this functional $\frac{F}{\rho_d}$ in Figure 6 for three representative values of $\varepsilon$ namely $\varepsilon=-\frac{1}{3},-\frac{2}{3}$ and $-1$. Baring for some cosmic time in the early phase, $\frac{F}{\rho_d}$ remains in the positive domain all through the cosmic evolution. It increases with time, peaks at around $t=1.5$ and then decreases with the cosmic dynamics. At large cosmic time, $\frac{F}{\rho_d}$ becomes time independent and asymptotically reduces to  $\frac{3(m+1)}{m^2+4m+1}$. For all the cases of bulk viscous barotropic cosmic fluid considered here, the functional $\frac{F}{\rho_d}$ behaves alike through out the cosmic evolution except near the peak. The peaks for different $\varepsilon$ are different. The peak is higher for $\varepsilon=-1$ and lower for $\varepsilon=-\frac{1}{3}$.

\begin{figure}[h!]
\begin{center}
\includegraphics[width=1\textwidth]{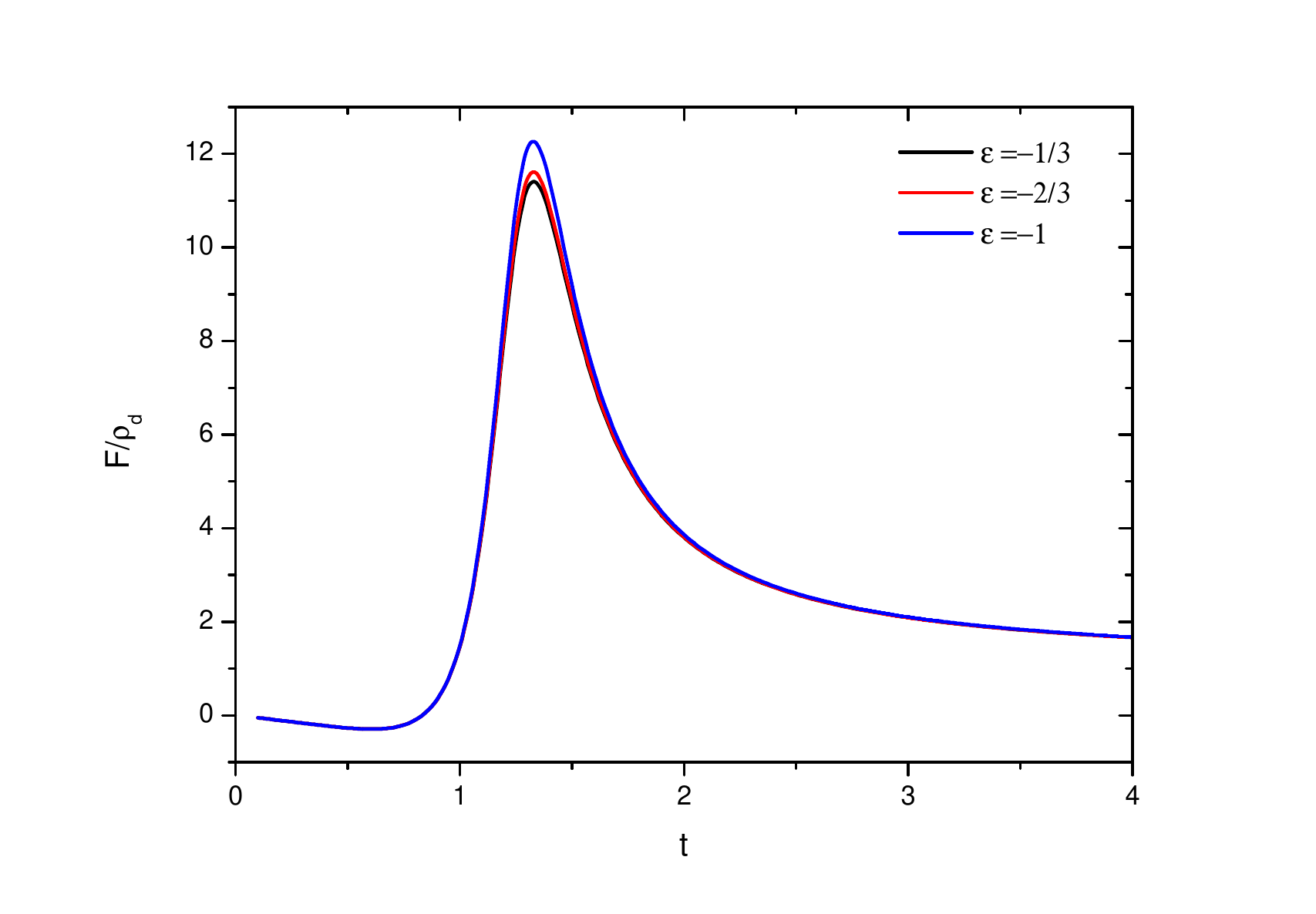}
\caption{Evolution of the functional $\frac{F(t)}{\rho_d}$  with cosmic time for the de Sitter model of expansion for three representative value of the parameter $\varepsilon$.}
\end{center}
\end{figure}

\begin{figure}[h!]
\begin{center}
\includegraphics[width=1\textwidth]{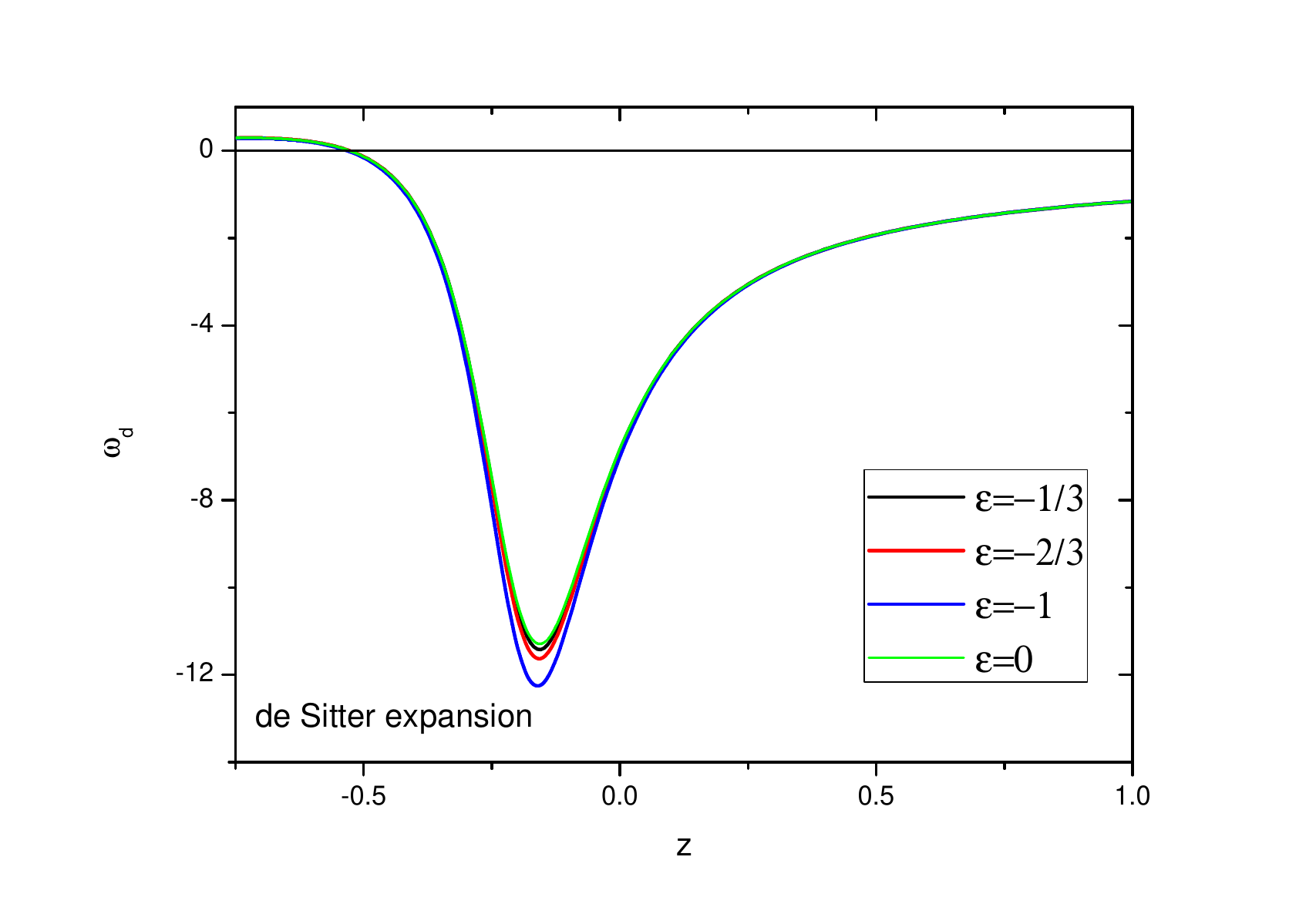}
\caption{Dynamics of the dark energy equation of state parameter  $\omega_d$ for de Sitter model. Four representative values of the parameter $\varepsilon$ are considered.}
\end{center}
\end{figure}

In Figure-7, $\omega_d$ is plotted for four different values of $\varepsilon$, namely $ \varepsilon=0,-\frac{1}{3}, -\frac{2}{3}$ and $-1$. The case $\varepsilon=0$ corresponds to the cosmic fluid with no usual matter but of only dark fluid.  The dark energy EoS, for all the cases considered here, decreases from some constant value at early cosmic phase becomes the lowest for some particular redshift $z=-0.16$ and then increases with cosmic time. At a redshift $z=-0.525$ it vanishes and becomes positive at some future cosmic time. The behavioural switching over of $\omega_d$ is predicted to occur at a future time. At the present epoch, it maintains the same decreasing trend and lies in the phantom region. The bulk viscous cosmic fluid has a very little impact on the general behaviour of the EoS parameter. For all the values of $\varepsilon$ considered in the work, it maintains the same evolutionary trend. However, near the well ( negative peak), the value of $\omega_d$ is the lowest for $\varepsilon= -1$  and highest for $\varepsilon=0$. 

$\omega_d$ is very much sensitive to the choice of $\xi$. In Figure 8, we have shown the EoS parameter for the present de Sitter model for different choices of $\xi$. With the increase in the value of $\xi$, the behavioural switching over time of the universe shifts towards the past. However, at very early time and at late cosmic time, $\omega_d$ becomes independent of the choice of $\xi$. In Figure 9 we have shown $\omega_d$ for different choices of $\alpha$. $\omega_d$ remains the same for all values of $\alpha$ both at early and late cosmic evolutionary phases but differs in the middle phase of cosmic evolution. With the increase in the value of $\alpha$, the decrement in $\omega_d$ is more rapid. Also, the value of $\omega_d$ near the well ( negative peak) is more for higher value of $\alpha$.

\begin{figure}[h!]
\begin{center}
\includegraphics[width=1\textwidth]{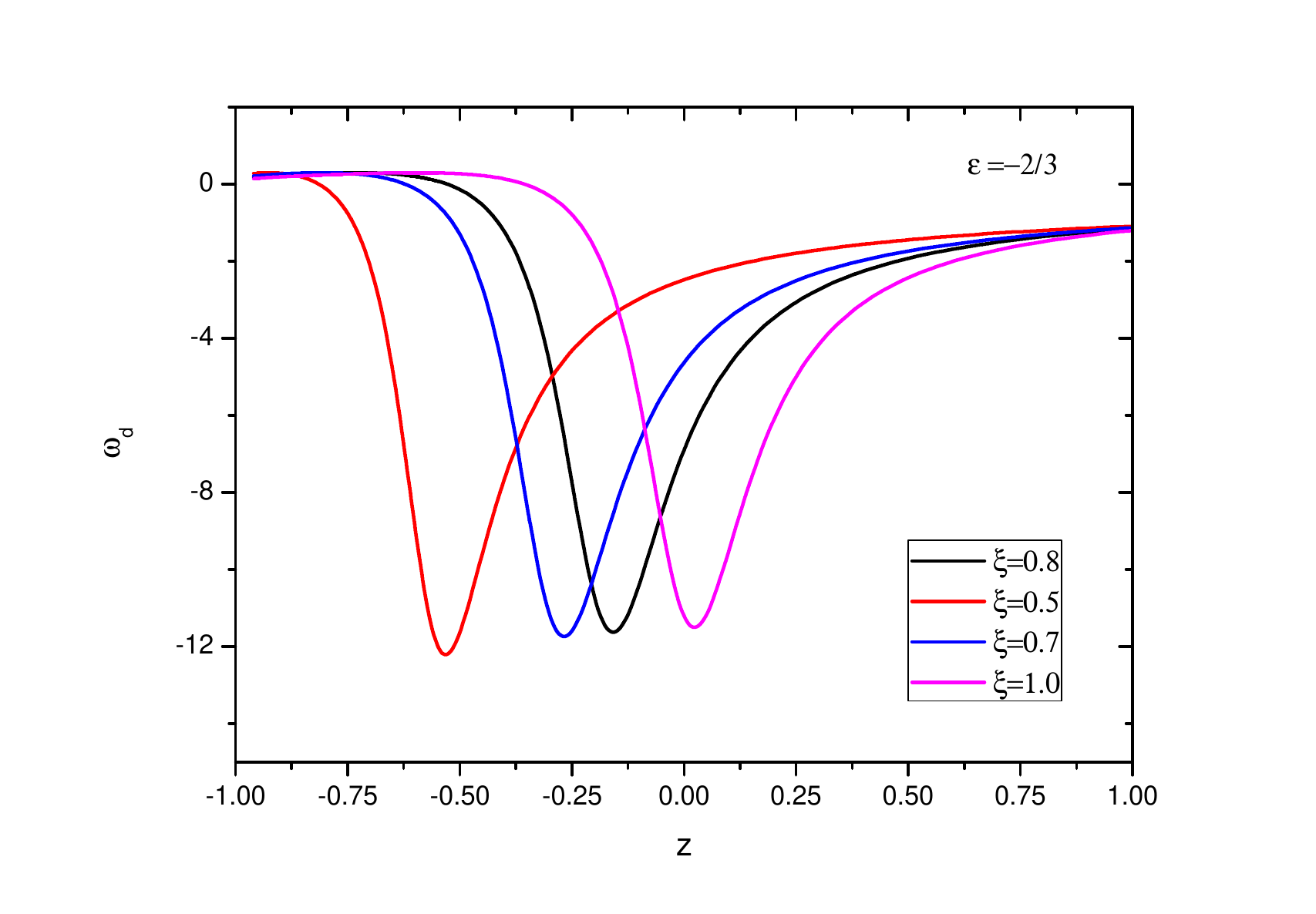}
\caption{Variation of $\omega_d$ with the parameter $\xi$ of the de Sitter expansion model.}
\end{center}
\end{figure}

\begin{figure}[h!]
\begin{center}
\includegraphics[width=1\textwidth]{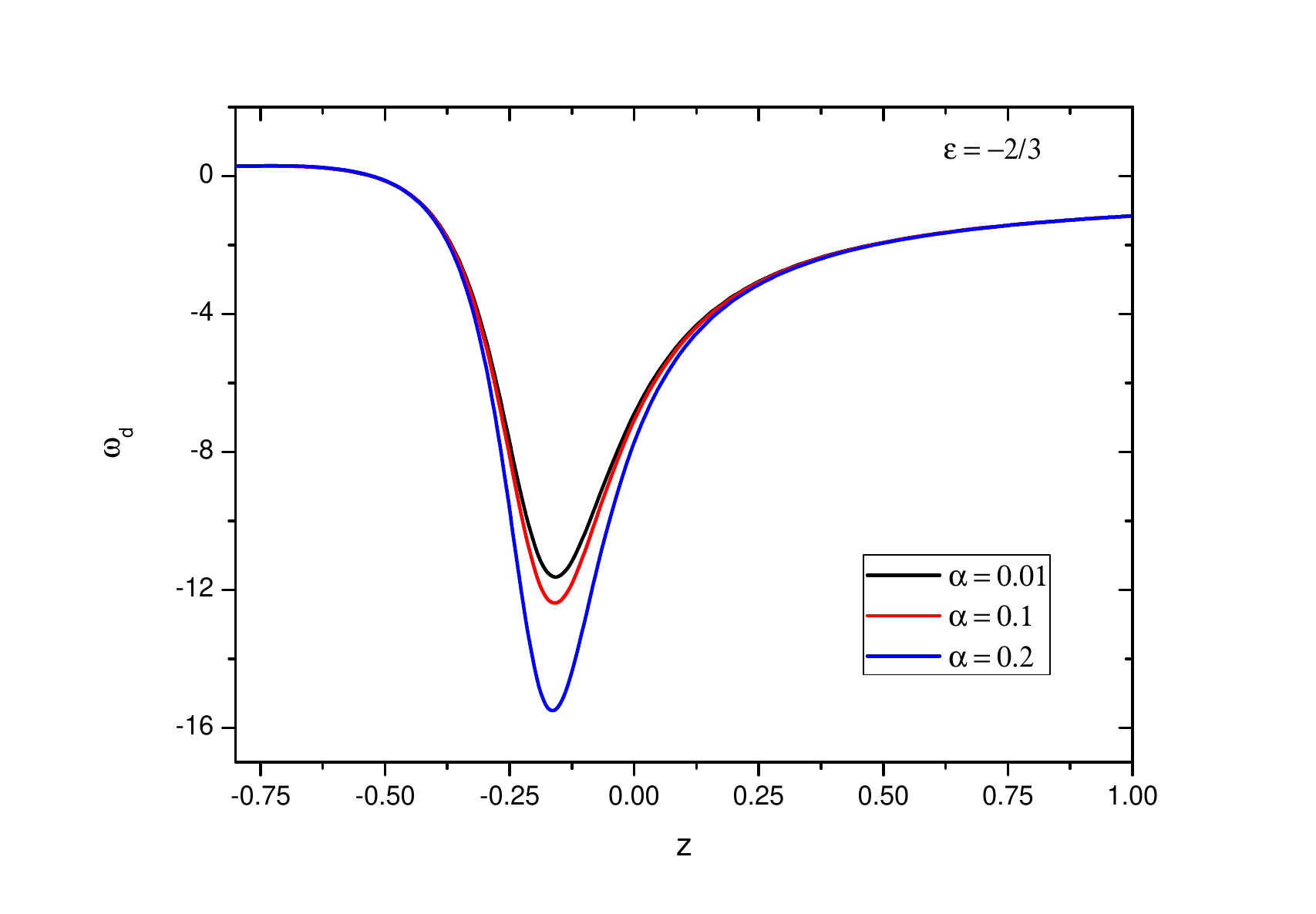}
\caption{Variation of $\omega_d$ with the parameter $\alpha$ of the anisotropic metric BV (refer to eqn. 5) for the de Sitter model of expansion for the case  $\varepsilon =-2/3$. }
\end{center}
\end{figure}

The evolution of skewness parameters with cosmic dynamics is shown in Figure 10. In the figure, the time evolution of the functional $\frac{F(t)}{\rho_d}$ for a particular choice of $\varepsilon=-\frac{2}{3}$ is also shown for reference. It can be noted from the figure that the behaviour of the skewness parameters depend on the behaviour of $\frac{F(t)}{\rho_d}$. At early cosmic phase $\eta$ starts with a negative value close to zero and switches over to positive  values at a redshift of $z=0.14$. $\eta$ increases with the cosmic time to become maximum at redshift $z=-0.23$ and then again decreases with time. The evolutionary behaviour of $\gamma$ is just the mirror image of $\eta$. The pressure anisotropy along the x-axis vanishes as expected earlier. It is interesting to note that, at the switching over redshift $z=0.14$, all the skewness parameter vanish.


\begin{figure}[h!]
\begin{center}
\includegraphics[width=1\textwidth]{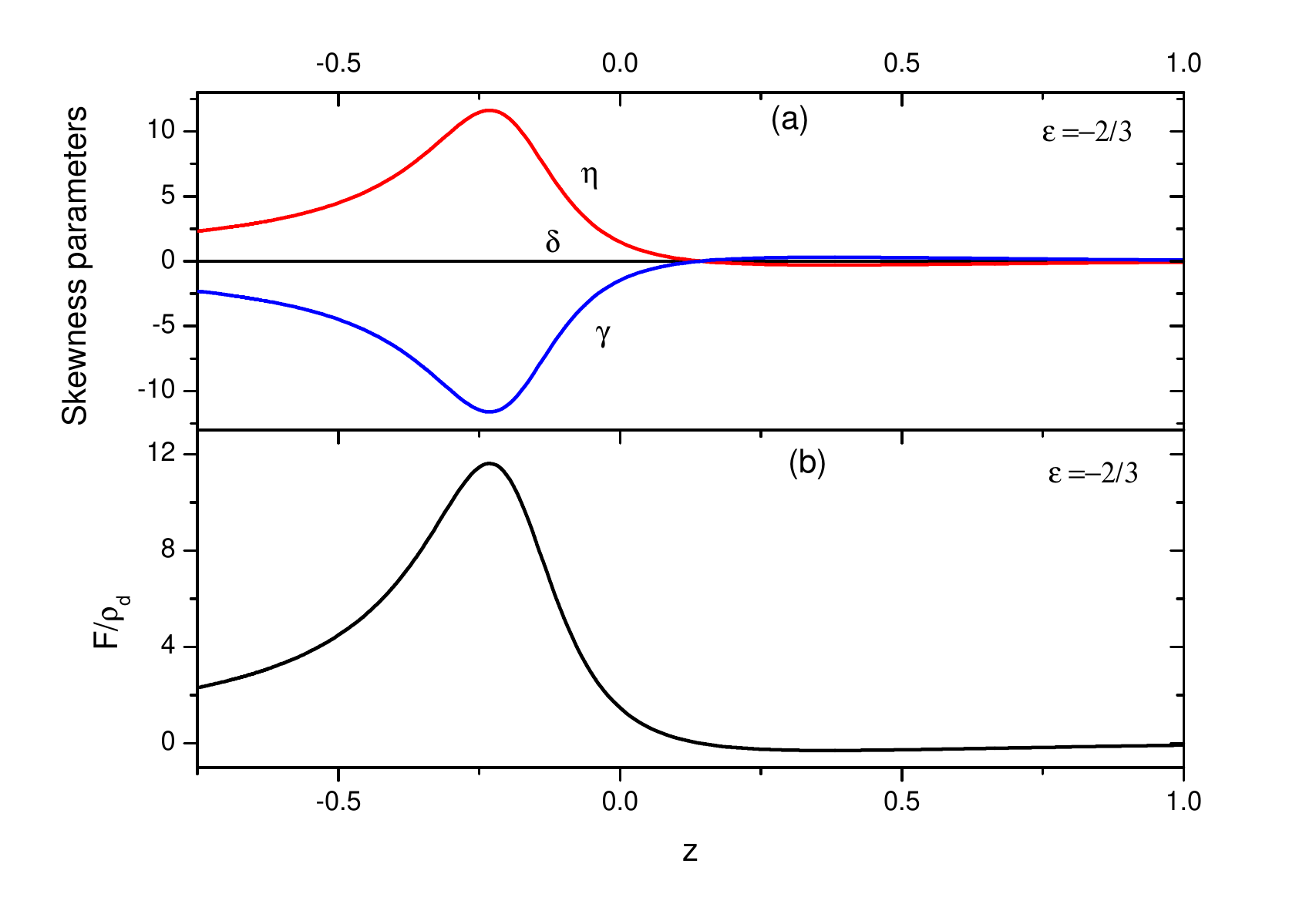}
\caption{(a)(upper panel) Dynamical evolution of skewness parameters along three spatial directions with respect to redshift for the de Sitter model for the case  $\varepsilon =-2/3$. (b)  (lower panel) Since the behaviour of  $\frac{F(t)}{\rho_d}$ controls the nature of the pressure anisotropies in dark energy fluid, its evolution  as function of redshift is shown for $\varepsilon =-2/3$ for reference. }
\end{center}
\end{figure}


\section{Summary and Conclusions}

In the present work, we reconstructed some cosmological models in the frame work of Wesson's scale invariant gravity theory. The scale invariant theory we have adopted has already been claimed by Wesson to be a simple one that has passed some tests and may be considered superior to others proposed earlier. The gauge function is considered to depend only on cosmic time in a reciprocal manner. We consider non interacting bulk viscous cosmic fluid and dark energy as the source of matter field. Pressure anisotropy is assumed along different directions.  Power law  and de Sitter expansion laws are considered in the work. From the reconstructed models, we have studied the dynamics of the universe through the dark energy EoS parameter. The skewness parameter or pressure anisotropies are also calculated. The role of a bulk viscous fluid in addition to dark components is also investigated.

We found that the skewness parameters evolve dynamically with time. Along the x-axis, pressure anisotropy is negligible. In the two other spatial directions, i.e. y- and z- direction, the pressure anisotropies behaves just like the mirror image of one another. In the power law model, the pressure anisotropies almost remain constant through out the cosmic evolution. However, at future phase, they tend to increase.  In the de Sitter model, at early times, the universe is predicted to have almost isotropic fluid which becomes anisotropic with time. At late phase, pressure anisotropy again tends to decrease. There remains a residual pressure anisotropy even at late phase. The presence of a non interacting bulk viscous cosmic fluid along with the dark fluid does not substantially affect the dynamics of pressure anisotropy.

The present model mostly favours a phantom phase with $\omega_d$ lying primarily below the phantom divide i.e.  $\omega_d < -1$. The dark energy EoS parameter becomes time independent for two specific choices of $\varepsilon$ namely $\varepsilon = 0$ and $-1/3$ in power law model. For $\varepsilon = 0$, the model behaves like a cosmological constant with $\omega_d = -1$. With a decrease in the value of $\varepsilon$ in the negative domain, the DE EoS decrease to acquire more phantom energy. In the de Sitter model, we get interesting results for the dynamic evolution of the DE EoS parameter. $\omega_d$ decreases initially and after attaining a negative peak at certain cosmic time it again increases to exit from a catastrophic situation. At late phase of time it becomes constant. It is interesting to note that, this situation only occurs when we considered a scale invariant theory of gravity but in the absence of scale invariance in the field equations, no such behaviour is seen. In the absence of scale invariance, the DE equation of state parameter decreases rapidly in the phantom region with $\omega_d < -1$. Bulk viscosity does not affect appreciably to the dynamics in the de Sitter model.

In the present work, we observed that the scale invariant theory used to reconstruct anisotropic dark  models, present more interesting results favouring phantom kind of behaviour and in conformity to recent observations. From our work, it is certain that pressure anisotropies in the dark energy fluid play some important and interesting roles in dark energy models.  However, more involved investigation of the role of the pressure anisotropy should be carried out for further understanding of the cosmic mechanism.


\section{Acknowledgement}
BM acknowledges the Extra Mural Research Funding  support of SERB-DST, New Delhi, India [No.- SR/S4/MS:815/13]. SKT and BM acknowledge the support of IUCAA, Pune(India) where a part of this work is carried out.


\begin{thebibliography}{99}

\bibitem{Riess98} A.G.Riess et al., Astron.J \textbf{116}, 1009 (1998).
\bibitem{Allen04} S.W. Allen et al., Mon.Not.Roy. Astron.Soc. \textbf{353}, 457 (2004).
\bibitem{Ein05} D.J. Eisenstein et al.( SDSS Collaboration), Astrophys. J \textbf{633}, 560 (2005).

\bibitem{Nojiri03}S. Nojiri and S.D.Odintsov, Phys. Rev. \textbf{D 68},123512(2003).
\bibitem{Harko11} T. Harko, F.S.N. Lobo, S. Nojiri, S.D. Odintsov, Phys. Rev. \textbf{D. 84}, 024020 (2011).
\bibitem{Harko10} T. Harko, F.S.N. Lobo, Eur. Phys. J. \textbf{C. 70}, 373 (2010).

\bibitem{Nojiri05} S. Nojiri, S.D.Odintsov and M. Sasaki, Phys. Rev. \textbf{D 71}, 123509 (2005).

\bibitem{Bamba12} K. Bamba, S. Capozziello, S. Nojiri, S. D. Odintsov, Astrophysics and Space Science \textbf{342}, 155 (2012).

\bibitem{Antoniou12} A. Antoniou and L. Perivolaropoulos, J. Cosmol. Astropart. Phys. \textbf{12}, 012 (2010).

\bibitem{Ade13a} P.A.R. Ade et al.( Planck Collaboration), Astron. Astrophys. \textbf{571}, A16(2014).

\bibitem{Campa06}L. Campanelli, P. Cea, L. Tedesco,  Phys. Rev. Lett. \textbf{97}, 131302 (2006).

\bibitem{Campa09} L. Campanelli,  Phys. Rev. \textbf{D 80}, 063006 (2009).

\bibitem{Gruppo07}A. Gruppuso,  Phys. Rev. \textbf{D 76}, 083010 (2010).

\bibitem{Akarsu10} O. Akarsu, C.B. Kilinc, Gen. Rel Grav. \textbf{42}, 119 (2010).

\bibitem{Campo12}S. del Campo, V. H. Cardenas, R. Herrera, Mod. Phys. Lett. \textbf{A 27}, 1250213(2012).

\bibitem{SKT15} B. Mishra, P.K. Sahoo, S.K. Tripathy,  Astrophys. Space Sci.  \textbf{356}, 163 (2015).

\bibitem {Wesson81a} P.S. Wesson, Astron. Astrophys.\textbf{102}, 45 (1981).

\bibitem {Wesson81b} P.S. Wesson, Mon. Not. R. Astro. Soc.\textbf{39}, 429 (1981).

\bibitem {Dirac74} P.A.M. Dirac, Proc. R. Soc. Lon. \textbf{A338}, 439(1974).

\bibitem {Canuto77} V. Canuto, P.J. Adams, S.H. Hsieh,E. Tsiang,  Phys. Rev. \textbf{D16}, 1643(1977).
\bibitem {Bees86} A. Beesham, Clasic. Quant. Grav. \textbf{3}, 481(1986).
\bibitem {Mohanty03} G. Mohanty, B. Mishra, Astrophys. Space Sci.\textbf{283}, 67(2003).

\bibitem {Mishra04} B. Mishra, Chinese. Phys. Lett.\textbf{21(12)}, 2359(2004).

\bibitem {Mishra12a} B. Mishra, P.K. Sahoo, Int. J. pure and Appl. Maths.\textbf{80(4)}, 535(2012a).

\bibitem {Mishra12b} B. Mishra, P.K.Sahoo, Int. J. Theo. Phys.\textbf{51(2)}, 399(2012b).

\bibitem {Mishra14}B. Mishra, P.K. Sahoo, Astrophys. Space Sci.\textbf{349}, 491(2014).
\bibitem{Barrow06} J.D.Barrow, T. Clifton, Clas. Quant. Gravit. \textbf{23}, L1(2006). 

\bibitem{Shapo09} M. Shaposhnikov, D. Zenhausern, Phys. Lett. \textbf{671}, 187 (2009).
\bibitem{Brevik17} I. Brevik, $\varnothing$. Gr$\varnothing$n, J. de Haro, S. D. Odintsov, E. N. Saridakis, arxiv: 1706.02543 (2017).

\bibitem{SKT10}S.K.Tripathy, D.Behera and T.R.Routray, Astrophys. Space Sci.\textbf{325}, 2193(2010).
\bibitem{Collins80}C.B.Collins, E.N.Glass, D.A.Wilkinson, Gen. Relativ. Gravit. \textbf{12}, 805 (1980).

\bibitem{BMSKT15} B. Mishra and S.K.Tripathy, Mod. Phys. Lett. A \textbf{30},1550175 (2015).

\bibitem{SKT09}S.K.Tripathy, S. K.Nayak, S. K. Sahu and T.R.Routray, Astrophys. Space Sci.\textbf{325}, 93(2010).
\bibitem {Campa11} L Campanelli et al., Int.Jou. Mod. Phys. \textbf{D 20}, 1153 (2011).

\bibitem{Bunn96} E. F. Bunn, P. G. Ferreira, J. Silk, Phys. Rev. Lett. \textbf{77}, 2883 (2016).

\bibitem{Saadeh16} D. Saadeh, S. M. Feeney, A. Pontzen, H. V. Peiris,J.D.McEwen, Phys. Rev. Lett. \textbf{117}, 131302 (2016).
\bibitem{Rapetti07} D.Rapetti, S.W. Allen,  M.A.Amin, R.D. Blanford,  Mon.Not.Roy. Astron.Soc. \textbf{375}, 1510 (2007) , arxiv: astro-ph 0605683.

\bibitem{Giostri12} R.Giostri et al., J. Cosmol. Astropart. Phys., \textbf{1203}, 027 (2012), arxiv: astro-ph 1203.3213.

\end{thebibliography}
\end{document}